# High-Mobility Bismuth-based Transparent P-Type Oxide from High-throughput Material Screening

Amit Bhatia[1,†], Geoffroy Hautier[2,†,*], Tan Nilgianskul[1], Anna Miglio[2], Gian-Marco Rignanese[2], Xavier Gonze[2], Jin Suntivich[1,*]

[1] *Department of Materials Science and Engineering,*
*Cornell University, Ithaca, New York 14853, USA*

[2] *Institute of Condensed Matter and Nanosciences (IMCN),*
*Université Catholique de Louvain, Louvain-la-Neuve 1348, Belgium*

[†] These authors contribute equally to this work

[*] Correspondence: geoffroy.hautier@uclouvain.be, jsuntivich@cornell.edu

ABSTRACT

Transparent oxides are essential building blocks to many technologies, ranging from components in transparent electronics[1,2], transparent conductors[3,4], to absorbers and protection layers in photovoltaics and photoelectrochemical devices[5,6]. However, thus far, it has been difficult to develop p-type oxides with wide band gap and high hole mobility; current state-of-art transparent p-type oxides have hole mobility in the range of < 10 cm$^2$/V·s[7,8], much lower than their n-type counterparts[9-11]. Using high-throughput computational screening to guide the discovery of novel oxides with wide band gap and high hole mobility, we report the computational identification and the experimental verification of a bismuth-based double-perovskite oxide that meets these requirements. Our identified candidate, $Ba_2BiTaO_6$, has an optical band gap larger than 4 eV and a Hall hole mobility above 30 cm$^2$/V·s. We rationalize this finding with molecular orbital intuitions; $Bi^{3+}$ with filled *s*-orbitals strongly overlap with the oxygen *p*, increasing the extent of the metal-oxygen covalency and effectively reducing the valence effective mass, while $Ta^{5+}$ forms a conduction band with low electronegativity, leading to a high band gap beyond the visible range. Our concerted theory-experiment effort points to the growing utility of a data-driven materials discovery and the combination of both informatics and chemical intuitions as a way to discover future technological materials.

INTRODUCTION

Successful implementations of many energy and transparent electronic applications, ranging from transparent complementary transistors[12,13], high-power electronics, to photovoltaics and solar fuel systems[14,15], hinge on the discovery of a oxide material with good carrier mobility and visible transparency. Many n-type oxides with good electron mobility and visible transparency, such as ZnO, and In-Sn-O and In-Ga-Zn-O materials[1,9,10,16], have stellar performance and are already in use in many devices. However, this stands in contrast to the case of the p-type, where the performance has struggled to achieve the same level as the n-type. This limitation is widely believed to be due to the localization of the oxygen *2p* state in the valence band[17,18]. Driven by this realization, Hosono and co-workers have postulated that the key to unlocking the high hole mobility is to delocalize the traditionally localized oxygen *2p* state, effectively decreasing the hole effective mass. They proposed an approach to delocalize the oxygen *2p* by incorporating a highly electronegative cation that has energy levels close to the oxygen *2p* to increase the extent of the metal-oxygen covalency[19]. This concept has led to the discovery of Cu-containing compounds as high-performance wide-band-gap, p-type oxides[17,20] and oxysulfides[21-25]. This result has to date become the benchmark for the high-band-gap hole mobility (~1-10 $cm^2/V \cdot s$). However, despite this discovery, the transparent p-type Cu-based compounds are still limited by their low hole mobility[23,24,26-30]. When comparing to what have been achieved for the n-type (>100 $cm^2/V \cdot s$)[9-11], it can be seen that the performances of the Cu-based compounds are still modest.

In an effort to find superior p-type compounds, researchers have investigated the more spatially extended *s*-orbital chemistry, which has been postulated and demonstrated computationally to delocalize the oxygen *2p* more effectively[7,31-33]. However, so far, the experimental realization of the *s*-orbital-based transparent oxide has been limited. Tin monoxide, SnO ($Sn^{2+}$: [Kr] $4d^{10}$ $5s^2$) has shown to be the best p-type *s*-orbital candidate, however, with still moderate hole mobility (< 5 cm$^2$/V·s), anisotropic transport, low average transmission (75-80%) and limited air stability[7,8,32,34,35]. $Bi^{3+}$ ([Xe] $4f^{14}$ $5d^{10}$ $6s^2$)-compounds offers an alternative approach to utilizing the *s*-orbital chemistry[33]. However, the Bi *6s* states in current Bi-based oxides are too low for effective O *2p* hybridizations[7]. To fully take advantage of the more spatially extended *s*-orbital chemistry such as $Bi^{3+}$, it is essential to find a structure and chemistry that can support and promote the Bi *6s* – O *2p* hybridization while retaining the visible transparency. We have recently reported the use of band gap and valence band curvatures as parameters for screening for high-figure-of-merit p-type oxides from the binary and ternary oxide databases[18]. In the following, we report the experimental realization of an *s*-orbital bismuth-based candidate with strong metal–oxygen *s*–*p* hybridization and visible transparency, emerging from our high-throughput computational screening. We present its synthesis, optical and electrical characterization and provide a preliminary demonstration of its high transparency and hall mobility, suggesting that this compound can serve as a candidate for a future p-type transparent oxide compound.

IDENTIFICATION OF THE CANDIDATE MATERIAL

Our candidate oxide, $Ba_2BiTaO_6$ (BBT), was identified with the three constraints: low valence band effective mass in all crystal directions (< 0.5 $m_e$), visible transparency (band gap > 3 eV), and high valence band maximum with respect to vacuum (hence favoring p-type dopability, see Supporting Information). BBT has a rhombohedral double perovskite oxide structure[36-38], as shown in Fig. 1A, containing alternating $Bi^{3+}$ and $Ta^{5+}$ ('B-site') octahedra, surrounded by $Ba^{2+}$ ('A-site'). The unique qualities of BBT are the low effective mass of its valence band, its high valence band maximum, which favors high hole mobility, and its large band gap, which gives the transparency in the visible range (Fig. 1B). The first quality stems from the mixed Bi *6s* – O *2p* state as illustrated by the atomic projection of the valence band maximum (Fig. 1B, bonding analysis is shown in the Supporting Information). The charge density associated with the valence band shows that the Bi *s*-orbital hybridizes with the oxygen p-orbital pointing toward the octahedron center while the two others perpendicular O *p*-orbitals stay non-bonding (Fig. 1C). The conduction band on the other hand is of a mixed Bi, Ta and O character and shows less dispersion (Fig. 1A-B). We note that the valence band nature in BBT is similar to the $BaBiO_3$ perovskite, which has been extensively studied for its superconducting properties[39-42]. However, different from $BaBiO_3$, which has an optical gap around 2 eV[43,44], BBT is transparent due to the presence of $Ta^{5+}$ (instead of $Bi^{5+}$), which pushes the conduction band upward, effectively increases the band gap and gives rise to the transparency. Interestingly, the electronic structure of BBT mimics exactly what Sleight called the 'Holy Grail' electronic structure for p-type oxide[33] with an *s*-based cation hybridizing with the oxygen. However, unlike other Bi-based oxides studied in the past, which generally have their Bi *6s* state below the O *2p*[7,30], effectively limiting the Bi *6s* – O *2p* mixing, our chosen candidate has the Bi *6s* level close the O *2p*, which allows for a stronger Bi *6s* – O *2p* hybridization. At present, we attribute this unique Bi *6s* position to the BBT structure, although future study would be required to unravel the physical origin of this Bi *6s* upshift. We note that the finding of a compound with this strong Bi *6s* – O

—5—

2p hybridization and visible transparency was made possible by our ability to screen thousands of oxide candidates; finding this unique chemistry would have taken much longer time by trials and errors.

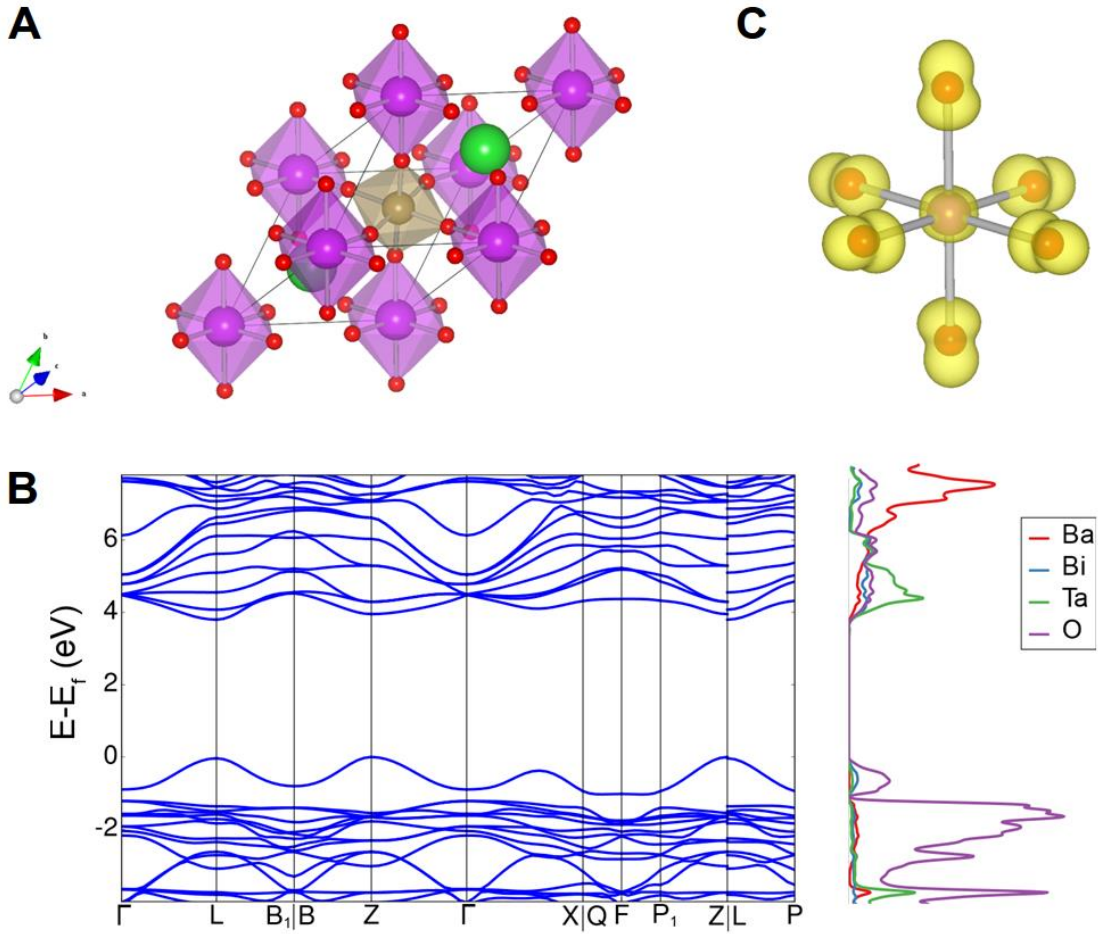

**Figure 1.** Crystal and electronic structure for $Ba_2BiTaO_6$. (A) Crystal structure of $Ba_2BiTaO_6$ in the rhombohedral space group (R-3m). (B) Band structure of $Ba_2BiTaO_6$ along symmetry lines and density of state projected on the different atomic sites using GGA. A scissor operator was applied to the band gap to make it fit the single-shot GW value. (C) Square of the wave function in DFT-GGA for the valence band in a $BiO_6$ octahedron.

EXPERIMENTAL CHARACTERIZATION

To experimentally verify the prediction, we synthesize a pure phase BBT using a solid-state synthesis method. The phase purity was verified with X-ray diffraction (XRD, Fig. 2A). The BBT powder sample was used as a pulsed laser deposition (PLD) target for BBT thin film deposition. Using (100) MgO as a substrate for growth, we obtained a textured BBT film (Fig. 2B-C) with a film thickness estimated to be ~120 nm using spectroscopic ellipsometry. The optical transmittance of the BBT film had an average value of > 90% in the wavelength range between the 350 – 800 nm (Fig. 3A). The direct optical band was estimated to be >4.5 eV (see Fig. 3B). The difference with the GW computed band gap (3.8 eV) could come from the difference between the optical and electronic gap, or from the possible gap underestimation present in the single-shot GW.

We find that the undoped BBT exhibited negligible conductivity in both the pellet and the thin film forms. We therefore examine the possibility of substituting an electron acceptor in BBT. In this study, we focus on using $K^+$ as a $Ba^{2+}$ aliovalent substitution, as $K^+$ has a similar ionic size as $Ba^{2+}$ and that $K^+$ was also previously used to substitute in place of Ba in $BaBiO_3$[39]. We found that $K^+$ can be substituted up to 35% of $Ba^{2+}$ in BBT (forming $Ba_{1.3}K_{0.7}BiTaO_6$, BKBT) with no noticeable secondary phase (Fig. 2A). Interestingly, the optical band gap was found to decrease with $K^+$ concentration (Fig. 3A-B), which could be from the reduction in the lattice parameter once $K^+$ substitutes in the $Ba^{2+}$ position. In any case, the visible transparency was largely preserved (> 90%), likely pointing to the lack of the mid-gap state formations with $K^+$ inclusion.

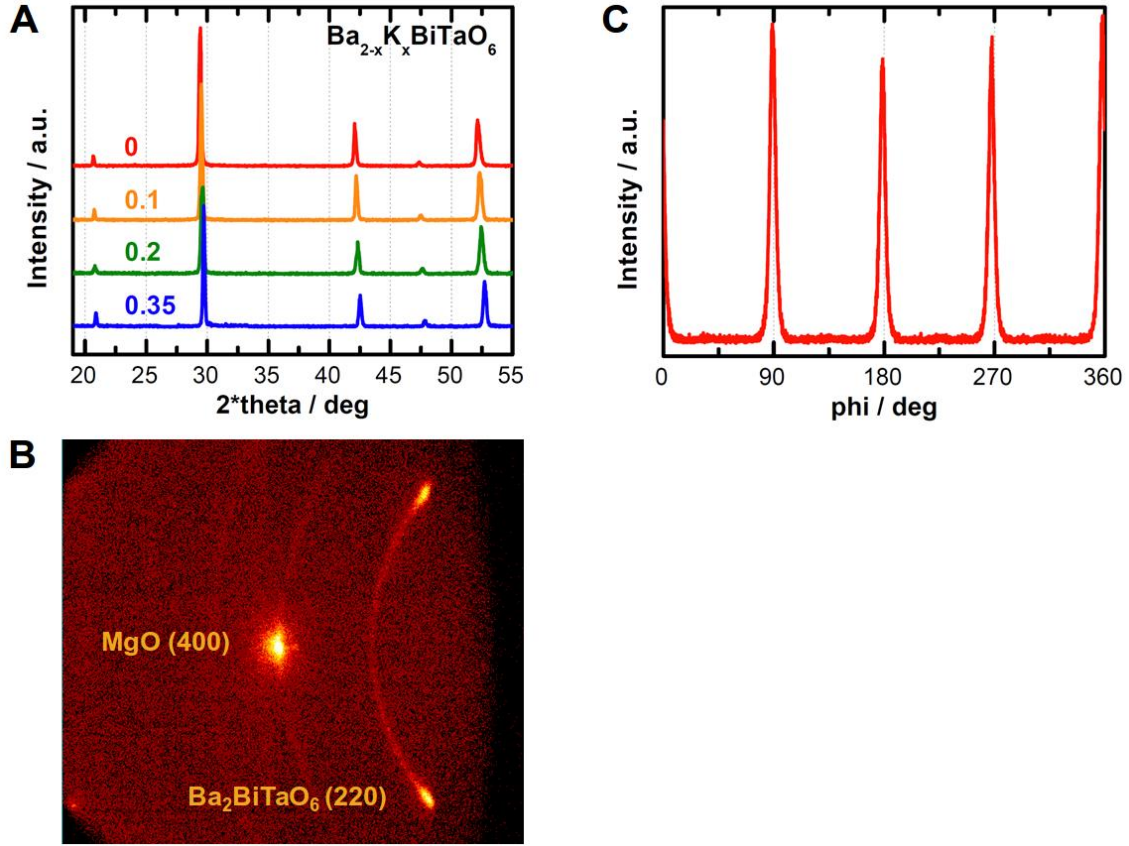

**Figure 2.** X-ray diffraction of the $Ba_2BiTaO_6$ powder and thin film. (A) Powder X-ray diffraction spectra for $Ba_{2-x}K_xBiTaO_6$. (B) Bruker General Area Detector Diffraction System (GADDS) of the $Ba_2BiTaO_6$ thin film. (C) Phi scan of the (220) peak of the $Ba_2BiTaO_6$ thin film. Both the GADDS and phi scans suggest that our $Ba_2BiTaO_6$ as grown on MgO is texturized.

Having demonstrated the visible transparency of BBT and K-doped BBT (BKBT), we now focus on evaluating the transport quality of the materials. Without K incorporation, BBT exhibits no detectable conductivity, behaving as an intrinsic wide-band-gap semiconductor. We thus focus on BKBT, specifically $Ba_{1.3}K_{0.7}BiTaO_6$, for the Hall experiment. Using the obtained Hall coefficient (see Supporting Information), we found the material to be p-type. We further extract the carrier concentration from the Hall experiment, which was found to be ~5 x $10^{13}$ cm$^{-3}$. This is an astonishingly low carrier concentration given the amount of K$^+$ in the material; this level of carrier concentration has only been previously reported in high quality single crystal works[45-47]. If the carrier concentration from the Hall measurement is correct, this observed low carrier concentration suggests that our oxide is fully compensated by the donor-like defects. We note that our difficulty in obtaining the p-type conductivity even at high K$^+$ concentration is well known in the Ba$_{1-x}$K$_x$BiO$_3$ literature, where n-type conductivity was still persistent even at 40% K substitution (Ba$_{0.58}$K$_{0.42}$BiO$_6$)[48,49], suggesting a prevalent nature of the compensating defects in the bismuth oxo-perovskites. We use the Hall carrier concentration to estimate the hole mobility, which we found to be in excess of 30 cm$^2$/V·s. This measured mobility is higher than the value measured previously for Ba$_{0.58}$K$_{0.42}$BiO$_6$, where the mobility was found to be 2 cm$^2$/V·s, likely as a result of the electron being the dominant carrier in the Hall measurement. Combining the mobility data with the optical gap, our BKBT compound represents the highest Hall hole mobility for a p-type transparent oxide to our knowledge.

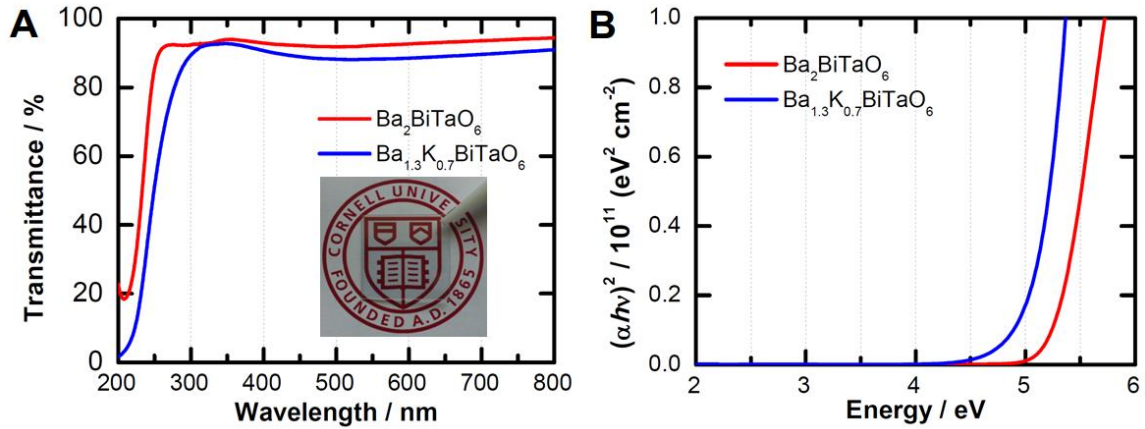

**Figure 3.** (A) Optical transmission and (B) $(\alpha h\upsilon)^2$ plot of the Ba$_{2-x}$K$_x$BiTaO$_6$ films.

We highlight that this preliminary value could be an underestimation of BKBT; we believe that significantly higher hole mobility can be attained with epitaxial films. Furthermore, it is important to recognize that our material is still not optimized and further studies will be required to understand and control its doping. The ionization ratio of our $K^+$ is nearly zero, which points to the formation of the 'hole-killing' defects that counter our $K^+$ incorporation. At this stage, it is difficult to know if BKBT will be able to reach the high carrier concentrations needed for "passive" transparent conducting oxides Nevertheless, the exceptional transmission and mobility properties justify a more careful attention at the very least for applications such as thin film transistor (TFT) for which the mobility could play an important role[1,16]. Efforts towards a full understanding of BKBT defects behavior and the development of a p-type channel TFT will be the subject of our future work.

CONCLUSION

In summary, we have reported on a new chemistry of high-mobility, visibly transparent p-type oxide: $Ba_2BiTaO_6$. We identified this oxide using high-throughput computational screening. Experimental characterizations show $Ba_2BiTaO_6$ to have >90% high transmission in the visible and the highest reported p-type mobility for a transparent oxide. To our knowledge, this is the first reported Bi *6s* p-type transparent compound, joining $Bi^{3+}$ to the selected club of elements leading to p-type transparent oxides (such as $Sn^{2+}$, $Cu^{1+}$ and $Ag^{1+}$). Although still limited by the carrier concentration, the exceptional performance of $Ba_2BiTaO_6$ is already evident in our preliminary characterizations. We further point out that $Ba_2BiTaO_6$ has a unique draw of having a ubiquitous perovskite structure, which offers a wide range of possibility of integrations with other functional perovskite oxide materials. Our work shows how the "needle in a haystack problem" of materials discovery can be accelerated using high-throughput material computations, and how material informatics can help guide the experimental realization of new technological materials.

METHODS

*Ab initio calculations.* The high-throughput screening was performed following the methodology reported previously[18,50]. Density functional theory (DFT) in the generalized gradient approximation (GGA)[51] with the VASP software[52] was used to assess hole effective masses. We used the materials project database[53,54] as our starting data set focusing on quaternary oxides. Follow-up computations using the more accurate single-shot GW method was performed to compute the material's band gap. GW computations were performed using ABINIT and norm-conserving pseudopotentials[55].

*Oxide Synthesis.* $Ba_{2-x}K_xBiTaO_6$ (with x ranging from 0 to 0.7) was prepared using a solid state synthesis method. Briefly, stoichiometric amounts of $BaCO_3$ (99.95%, Alfa Aesar), $Bi_2O_3$ (99.999%, Alfa Aesar) and $Ta_2O_5$ (99.993%, Alfa Aesar) were thoroughly ground with agate mortar and pestle. Then, the mixture was pressed into a pellet using a hydraulic press (Grimco) and heat treated in a furnace at 1100 °C for 12 hours under flowing argon (Ultrahigh purity, Airgas) atmosphere.

*Thin Film Deposition.* Pulsed laser deposition (PLD, Neocera) was used to deposit thin films of $Ba_{2-x}K_xBiTaO_6$ onto (100) MgO single-crystalline substrates (MTI). A pulsed laser beam from a KrF excimer laser ($\lambda = 248$ nm) was focused onto the target that was formed from the re-ground and re-pressed the oxide powders. The fluence energy and repetition rate of the laser were adjusted to achieve a deposition rate of ~ 5 nm/min. The substrate temperature and oxygen pressure were maintained at 650°C and 7.5 x $10^{-4}$ torr. After the deposition, the films were annealed at 400°C for 5 hours in an oxygen pressure of 1 torr.

*Structural and Optical Characterizations*. Crystal structures of the powder samples and films were characterized by X-ray Diffraction (Scintag Theta-Theta X-ray Diffractometer, Bruker General Area Detector Diffraction System, and Rigaku SmartLab X-ray Diffractometer) using Cu Kα radiation ($\lambda$ = 1.54052 Å). Optical transmission measurements were performed using UV-Vis-NIR Spectrometer (Shimazdu).

*Transport Measurement*. Powder samples of BKBT were pressed into pellets of 1 mm thickness, which was then cut into square bar specimens (7 x 7 mm) using diamond saw. The conductivity was measured using van der Pauw technique on contacts made from Ag-ink covered Ga-In eutectics (Sigma-Aldrich). Hole carrier density ($n_p$) and Hall mobility ($\mu_H$) were determined using Physical Property Measurement System (PPMS, Quantum Design) with Au/Ti electrodes at 300K.

ACKNOWLEDGEMENT

The authors would like to thank Dr. Christopher C. Evans and Chengyu Liu for their helps with ellipsometry, and Dr. Yuefeng Nie for the help with the Hall measurement. G.H. and G.M.R. acknowledge the F.R.S.-FNRS for financial support. The computational component of the work was supported by the European Union Marie Curie Career Integration (CIG) grant HTforTCOs PCIG11-GA-2012-321988. Computational resources have been provided by the supercomputing facilities of the Université catholique de Louvain (CISM/UCL) and the Consortium des Équipements de Calcul Intensif en Fédération Wallonie Bruxelles (CÉCI) funded by the Fond de la Recherche Scientifique de Belgique (FRS-FNRS). The experimental part of the work made use of the Cornell Center for Materials Research Shared Facilities which are supported through the NSF MRSEC program (DMR-1120296).

REFERENCES

1. Nomura, K. et al., Thin-film transistor fabricated in single-crystalline transparent oxide semiconductor. *Science* **300,** 1269-1272 (2003).
2. Morkoc, H. et al., Large-band-gap SiC, III-V nitride, and II-VI ZnSe-based semiconductor device technologies. *J. Appl. Phys.* **76,** 1363-1398 (1994).
3. Ginley, D. S., Hosono, H., & Paine, D. C., *Handbook of Transparent Conductors*. (Springer-Verlag Berlin, Berlin, 2010).
4. Ellmer, K., Past achievements and future challenges in the development of optically transparent electrodes. *Nat. Photonics* **6,** 808-816 (2012).
5. Izaki, M. et al., Electrochemically constructed p-$Cu_2O$/n-ZnO heterojunction diode for photovoltaic device. *J. Phys. D Appl. Phys.* **40,** 3326-3329 (2007).
6. Hu, S. et al., Amorphous $TiO_2$ coatings stabilize Si, GaAs, and GaP photoanodes for efficient water oxidation. *Science* **344,** 1005-1009 (2014).

7. Ogo, Y. et al., p-channel thin-film transistor using p-type oxide semiconductor, SnO. *Appl. Phys. Lett.* **93,** (2008).

8. Fortunato, E. et al., Transparent p-type $SnO_x$ thin film transistors produced by reactive rf magnetron sputtering followed by low temperature annealing. *Appl. Phys. Lett.* **97,** (2010).

9. Ohta, H. et al., Highly electrically conductive indium-tin-oxide thin films epitaxially grown on yttria-stabilized zirconia (100) by pulsed-laser deposition. *Appl. Phys. Lett.* **76,** 2740-2742 (2000).

10. Look, D. C., Droubay, T. C., & Chambers, S. A., Stable highly conductive ZnO via reduction of Zn vacancies. *Appl. Phys. Lett.* **101,** (2012).

11. Carcia, P. F., McLean, R. S., Reilly, M. H., & Nunes, G., Transparent ZnO thin-film transistor fabricated by rf magnetron sputtering. *Appl. Phys. Lett.* **82,** 1117-1119 (2003).

12. Kudo, A. et al., Fabrication of transparent p-n heterojunction thin film diodes based entirely on oxide semiconductors. *Appl. Phys. Lett.* **75,** 2851-2853 (1999).

13. Ohta, H. et al., Fabrication and photoresponse of a pn-heterojunction diode composed of transparent oxide semiconductors, p-NiO and n-ZnO. *Appl. Phys. Lett.* **83,** 1029-1031 (2003).

14. Stauber, R. E., Perkins, J. D., Parilla, P. A., & Ginley, D. S., Thin film growth of transparent p-type $CuAlO_2$. *Electrochem. Solid State Lett.* **2,** 654-656 (1999).

15. Tonooka, K., Bando, H., & Aiura, Y., Photovoltaic effect observed in transparent p-n heterojunctions based on oxide semiconductors. *Thin Solid Films* **445,** 327-331 (2003).

16. Nomura, K. et al., Room-temperature fabrication of transparent flexible thin-film transistors using amorphous oxide semiconductors. *Nature* **432,** 488-492 (2004).

17. Kawazoe, H., Yanagi, H., Ueda, K., & Hosono, H., Transparent p-type conducting oxides: Design and fabrication of p-n heterojunctions. *MRS Bull.* **25,** 28-36 (2000).

18. Hautier, G., Miglio, A., Ceder, G., Rignanese, G. M., & Gonze, X., Identification and design principles of low hole effective mass p-type transparent conducting oxides. *Nat. Commun.* **4,** (2013).

19. Suntivich, J. et al., Estimating Hybridization of Transition Metal and Oxygen States in Perovskites, from O K-edge X-ray Absorption Spectroscopy. *J. Phys. Chem. C* **118,** 1856-1863 (2014).

20. Kawazoe, H. et al., P-type electrical conduction in transparent thin films of $CuAlO_2$. *Nature* **389,** 939-942 (1997).

21. Scanlon, D. O., Buckeridge, J., Catlow, C. R. A., & Watson, G. W., Understanding doping anomalies in degenerate p-type semiconductor LaCuOSe. *J. Mater. Chem. C* **2,** 3429-3438 (2014).

22. Ueda, K., Inoue, S., Hirose, S., Kawazoe, H., & Hosono, H., Transparent p-type semiconductor: LaCuOS layered oxysulfide. *Appl. Phys. Lett.* **77,** 2701-2703 (2000).

23. Hiramatsu, H. et al., Degenerate p-type conductivity in wide-gap $LaCuOS_{1-x}Se_x$ (x = 0 - 1) epitaxial films. *Appl. Phys. Lett.* **82,** 1048-1050 (2003).

24. Ueda, K., Inoue, S., Hosono, H., Sarukura, N., & Hirano, M., Room-temperature excitons in wide-gap layered-oxysulfide semiconductor: LaCuOS. *Appl. Phys. Lett.* **78,** 2333-2335 (2001).

25. Scanlon, D. O. & Watson, G. W., $(Cu_2S_2)(Sr_3Sc_2O_5)$-A Layered, Direct Band Gap, p-Type Transparent Conducting Oxychalcogenide: A Theoretical Analysis. *Chem. Mat.* **21,** 5435-5442 (2009).

26. Pellicer-Porres, J. et al., On the band gap of $CuAlO_2$ delafossite. *Appl. Phys. Lett.* **88,** (2006).

27. Yanagi, H. et al., Electronic structure and optoelectronic properties of transparent p-type conducting $CuAlO_2$. *J. Appl. Phys.* **88,** 4159-4163 (2000).

28. Tate, J. et al., Origin of p-type conduction in single-crystal $CuAlO_2$. *Phys. Rev. B* **80,** (2009).

29. Kudo, A., Yanagi, H., Hosono, H., & Kawazoe, H., $SrCu_2O_2$: A p-type conductive oxide with wide band gap. *Appl. Phys. Lett.* **73,** 220-222 (1998).

30. Hiramatsu, H. et al., Crystal structures, optoelectronic properties, and electronic structures of layered oxychalcogenides MCuOCh (M = Bi, La; Ch = S, Se, Te): Effects of electronic configurations of $M^{3+}$ ions. *Chem. Mat.* **20,** 326-334 (2008).

31. Walsh, A., Payne, D. J., Egdell, R. G., & Watson, G. W., Stereochemistry of post-transition metal oxides: revision of the classical lone pair model. *Chem. Soc. Rev.* **40,** 4455-4463 (2011).

32. Ogo, Y. et al., Tin monoxide as an s-orbital-based p-type oxide semiconductor: Electronic structures and TFT application. *Phys. Status Solidi A-Appl. Mat.* **206,** 2187-2191 (2009).

33. Sleight, A., in *Handbook of Transparent Conductors*, edited by D. Ginley, H. Hosono, & D.C. Paine (Springer, 2011), pp. 295-311.

34. Quackenbush, N. F. et al., Origin of the Bipolar Doping Behavior of SnO from X-ray Spectroscopy and Density Functional Theory. *Chem. Mat.* **25,** 3114-3123 (2013).

35. Allen, J. P., Scanlon, D. O., Piper, L. F. J., & Watson, G. W., Understanding the defect chemistry of tin monoxide. *J. Mater. Chem. C* **1,** 8194-8208 (2013).

36. Wang, H. et al., Synthesis, Structure, and Characterization of the Series $BaBi_{1-x}Ta_xO_3$ ($0 \leq x \leq 0.5$). *Inorg. Chem.* **49,** 5262-5270 (2010).

37. Wallwork, K. S., Kennedy, B. J., Zhou, Q. D., Lee, Y., & Vogt, T., Pressure and temperature-dependent structural studies of $Ba_2BiTaO_6$. *J. Solid State Chem.* **178,** 207-211 (2005).

38. Zhou, Q. D. & Kennedy, B. J., High temperature structural studies of $Ba_2BiTaO_6$. *Solid State Sci.* **7,** 287-291 (2005).

39. Baumert, B. A., Barium potassium bismuth oxide: A review. *J. Supercond.* **8,** 175-181 (1995).

40. Hinks, D. G. et al., Synthesis, structure and superconductivity in the $Ba_{1-x}K_xBiO_{3-y}$ system. *Nature* **333,** 836-838 (1988).

41. Schneemeyer, L. F. et al., Growth and structural characterization of superconducting $Ba_{1-x}K_xBiO_3$ single crystals. *Nature* **335,** 421-423 (1988).

42. Tseng, D. & Ruckenstein, E., Structure and superconductivity of $BaBiO_3$ doped with alkali ions. *J. Mater. Res.* **5,** 742-745 (1990).

43. Wang, Y. Y., Feng, G. F., Sutto, T. E., & Shao, Z. F., Dielectric function of $BaBiO_3$ investigated by electron-energy-loss spectroscopy and ellipsometry. *Phys. Rev. B* **44,** 7098-7101 (1991).

44. Federici, J. F., Greene, B. I., Hartford, E. H., & Hellman, E. S., Optical characterization of excited states in BaBiO$_3$. *Phys. Rev. B* **42,** 923-926 (1990).

45. Utsch, B. & Hausmann, A., Halleffekt und Leitfähigkeitsmessungen an Zinkoxid-Einkristallen mit Sauerstofflücken als Donatoren. *Z Physik B* **21,** 27-31 (1975).

46. Ellmer, K., Resistivity of polycrystalline zinc oxide films: current status and physical limit. *J. Phys. D-Appl. Phys.* **34,** 3097-3108 (2001).

47. Fleischer, M. & Meixner, H., Electron mobility in single-crystalline and polycrystalline Ga$_2$O$_3$. *J. Appl. Phys.* **74,** 300-305 (1993).

48. Minami, H. & Uwe, H., Electrical conductivity of the oxide superconductor Ba$_{0.58}$K$_{0.42}$BiO$_{2.96}$. *J. Phys. Soc. Jpn.* **66,** 1771-1775 (1997).

49. Lee, S. F. et al., Hall effect and resistivity in metallic Ba$_{1-x}$K$_x$BiO$_3$ single crystals: Absence of 1/T dependence in R$_h$ and linear-to-quadratic evolution of ρ(T)*. *Physica C* **209,** 141-144 (1993).

50. Hautier, G., Miglio, A., Waroquiers, D., Rignanese, G. M., & Gonze, X., How Does Chemistry Influence Electron Effective Mass in Oxides? A High-Throughput Computational Analysis. *Chem. Mat.* **26,** 5447-5458 (2014).

51. Perdew, J. P., Burke, K., & Ernzerhof, M., Generalized gradient approximation made simple. *Phys. Rev. Lett.* **77,** 3865-3868 (1996).

52. Kresse, G. & Furthmuller, J., Efficient iterative schemes for ab initio total-energy calculations using a plane-wave basis set. *Phys. Rev. B* **54,** 11169-11186 (1996).

53. Jain, A. et al., Commentary: The Materials Project: A materials genome approach to accelerating materials innovation. *APL Mater.* **1,** (2013).

54. Ong, S. P. et al., Python Materials Genomics (pymatgen): A robust, open-source python library for materials analysis. *Comput. Mater. Sci.* **68,** 314-319 (2013).

55. Gonze, X. et al., ABINIT: First-principles approach to material and nanosystem properties. *Comput. Phys. Commun.* **180,** 2582-2615 (2009).

# Supporting Information

**S1. Valence band level vs vacuum**

To assess the typical nature of the doping, we performed a slab computation on the (100) orientation of the rhombohedral $Ba_2BiTaO_6$ (BBT) structure. We used to construct the slab the procedure presented in Sun *et al.*[S1]. The BBT slab computation was performed within GGA-PBE and we added the valence and conduction band correction from single-shot GW. Valence and conduction band edges can be used to assess dopability[S2]. Comparing BBT's valence band edge to typical p-type TCOs (e.g., $CuAlO_2$ and SnO) and n-type TCOs (e.g., $In_2O_3$ and ZnO), BBT is expected to behave more as a p-type than n-type material. Computations on a (110) slab gave similar results.

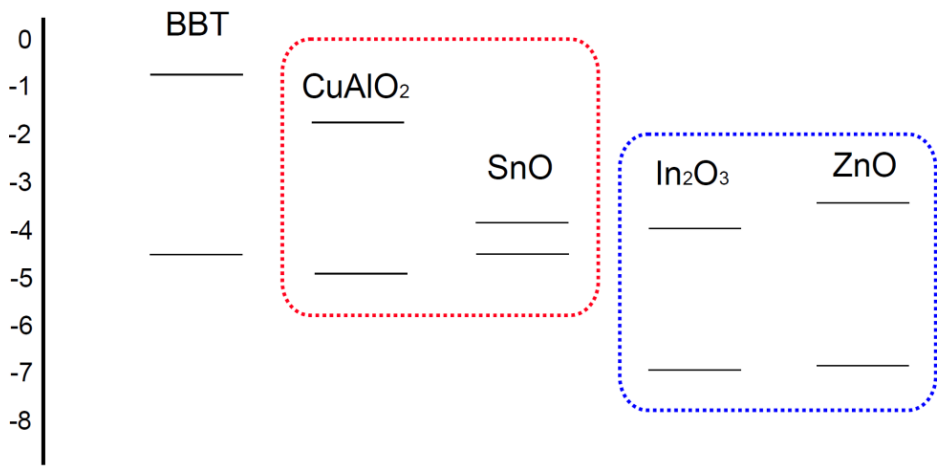

**Figure S1**. Conduction band and valence band levels vs vacuum for BBT compared to typical n-type TCOs ($In_2O_3$ and ZnO) as well as p-type TCOs ($CuAlO_2$ and SnO).

## S2. Orbital overlap analysis of the rhombohedral Ba$_2$BiTaO$_6$ structure

We performed an orbital overlap analysis in the COHP framework using the lobster software[S3-5] on our VASP GGA-PBE results. Figure S2 shows that the valence band is formed by anti-bonding between Bi *6s* and O *2p* states.

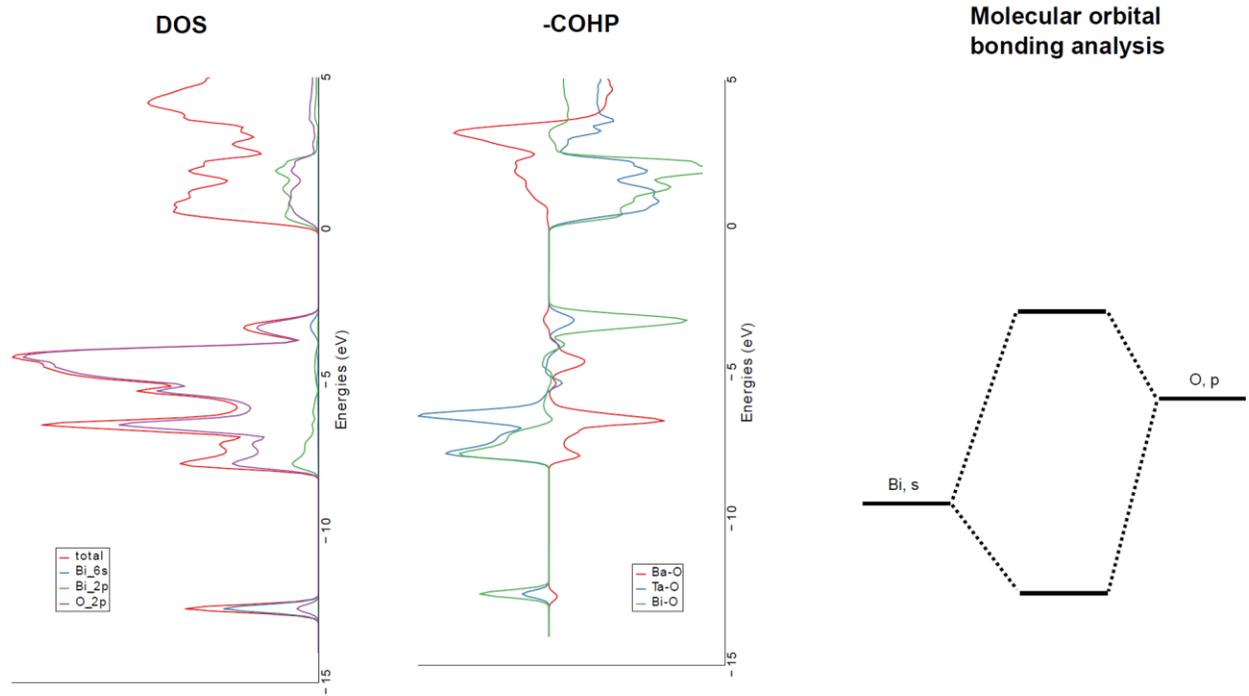

**Figure S2**. (Left) Density of state from a projection on localized basis sets using the lobster software on GGA-PBE results, (center) COHP analysis, and (right) schematic molecular orbital diagram for the Bi *6s* – O *2p* hybridization.

## S3. Indirect optical transition in $Ba_{2-x}K_xBiTaO_6$

We observe a systematic reduction in the BKBT band gap (both direct, Fig. 3B, and indirect) with increasing K. At the moment, the origin of this decrease is unclear.

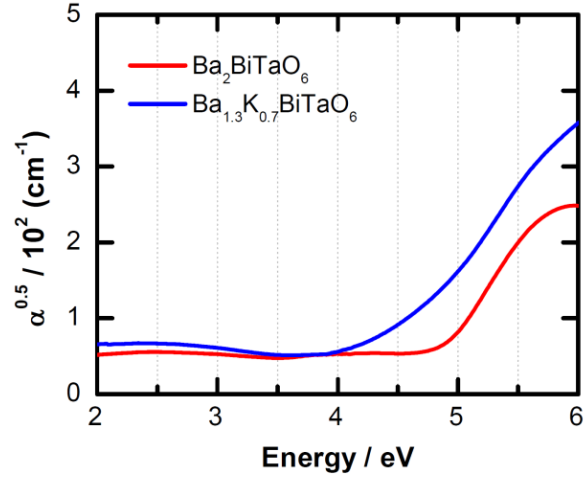

**Figure S3**. Plot of a square root of an absorption coefficient with photon energy, showing a decreasing indirect optical transition with $K^+$ substitution.

## S4. Hall measurement on $Ba_{1.3}K_{0.7}BiTaO_6$

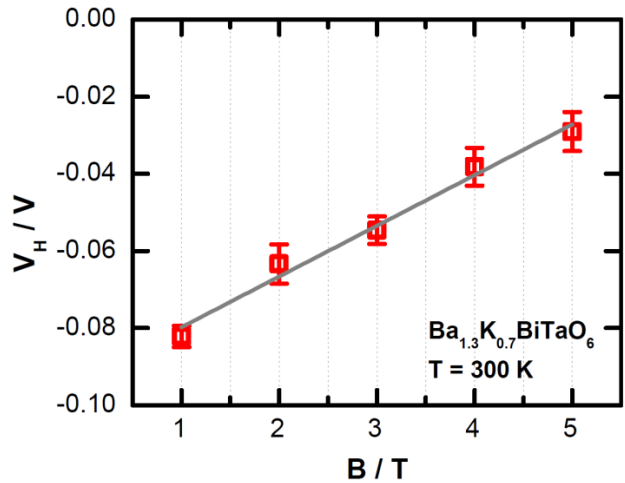

**Figure S4**. Hall voltages as a function of magnetic field. The slope was used to estimate the carrier concentration, which was calculated to be ~5 x $10^{13}$ $cm^{-3}$.

## S5. Additional References

S1.  Sun, W. H. & Ceder, G., Efficient creation and convergence of surface slabs. *Surf. Sci.* **617,** 53-59 (2013).

S2.  Robertson, J. & Clark, S. J., Limits to doping in oxides. *Phys. Rev. B* **83,** (2011).

S3.  Dronskowski, R. & Blochl, P. E., Crystal orbital Hamilton populations (COHP): energy-resolved visualization of chemical bonding in solids based on density-functional calculations. *J. Phys. Chem.* **97,** 8617-8624 (1993).

S4.  Deringer, V. L., Tchougreeff, A. L., & Dronskowski, R., Crystal Orbital Hamilton Population (COHP) Analysis As Projected from Plane-Wave Basis Sets. *J. Phys. Chem. A* **115,** 5461-5466 (2011).

S5.  Maintz, S., Deringer, V. L., Tchougreeff, A. L., & Dronskowski, R., Analytic Projection From Plane-Wave and PAW Wavefunctions and Application to Chemical-Bonding Analysis in Solids. *J. Comput. Chem.* **34,** 2557-2567 (2013).